\documentclass[10pt,conf]{IEEEtran}
\ifCLASSINFOpdf \else \fi
\usepackage{geometry}
\usepackage{multirow}
\usepackage{bbold}
\usepackage{ifpdf}
\usepackage[normalem]{ulem} 
\usepackage{subcaption}%
\usepackage[cmex10]{amsmath}
\usepackage{amssymb}
\usepackage{verbatim}
\usepackage{algorithm}
\usepackage{algorithmic}
\usepackage{array}
\usepackage{latexsym}
\usepackage{color}
\usepackage{textgreek}
\usepackage{subscript}
\usepackage{rotating}
\usepackage{booktabs,multirow}
\usepackage{mdframed}	
\usepackage{tabularx}
\usepackage{amsmath}
\usepackage{makecell}
\usepackage{tabto}
\usepackage{mathrsfs}
\usepackage{url}
\usepackage{cite}
\usepackage{listings}
\usepackage{array}
\usepackage[table]{xcolor}
\usepackage{ragged2e}
\usepackage{booktabs}

\usepackage{graphicx}
\usepackage{comment}
\usepackage[cmex10]{amsmath}
\usepackage{amssymb}
\usepackage{bbold}
\usepackage{float}
\usepackage[cmex10]{amsmath}
\usepackage{amssymb}
\usepackage{verbatim}
\usepackage{array}
\usepackage{latexsym}
\usepackage{color}
\usepackage{tabularx} 
\usepackage{textgreek}
\usepackage{subscript}
\usepackage{rotating}
\usepackage{booktabs,multirow}
\usepackage{mdframed}	
\usepackage{tabularx}
\usepackage{amsmath}
\usepackage{makecell}
\usepackage{tabto}
\usepackage{mathrsfs}
\usepackage{url}
\usepackage{cite}
\usepackage{listings}
\usepackage{array}
\usepackage[table]{xcolor}

\definecolor{pinkpurple}{rgb}{0.6, 0.1, 0.9} 
\usepackage{algorithmic}
\usepackage{array}
\usepackage{soul}
\sethlcolor{green}
\usepackage{url}
\usepackage{pifont}
\usepackage{xcolor}
\usepackage{multirow}
\usepackage{graphicx} 

\begin{document}

\title{Proactive SFC Provisioning with Forecast-Driven DRL in Data Centers}


\author{
\IEEEauthorblockN{Parisa Fard Moshiri~\IEEEmembership{Student Member, IEEE}, Poonam Lohan~\IEEEmembership{SMIEEE},\\ Burak Kantarci~\IEEEmembership{SMIEEE},~and~Emil Janulewicz\vspace{-0.1in}}\\
\thanks{
 P. Fard Moshiri, P. Lohan and B. Kantarci are with the School of Electrical Engineering and Computer Science, University of Ottawa, Ottawa, ON, Canada. Emails: \{ parisa.fard.moshiri,ppoonam,burak.kantarci\}@uottawa.ca

Emil Janulewicz is with Ciena, 383 Terry Fox Dr, Kanata, ON K2K 2P5, Canada, Email: ejanulew@ciena.com 
}
}

\maketitle
\thispagestyle{empty}
\pagestyle{empty}
\begin{abstract}

Service Function Chaining (SFC) requires efficient placement of Virtual Network Functions (VNFs) to satisfy diverse service requirements while maintaining high resource utilization in Data Centers (DCs). Conventional static resource allocation often leads to overprovisioning or underprovisioning due to the dynamic nature of traffic loads and application demands. To address this challenge, we propose a hybrid forecast–driven Deep reinforcement learning (DRL) framework that combines predictive intelligence with SFC provisioning. Specifically, we leverage DRL to generate datasets capturing DC resource utilization and service demands, which are then used to train deep learning forecasting models. Using Optuna-based hyperparameter optimization, the best-performing models, Spatio-Temporal Graph Neural Network, Temporal Graph Neural Network, and Long Short-Term Memory, are combined into an ensemble to enhance stability and accuracy. The ensemble predictions are integrated into the DC selection process, enabling proactive placement decisions that consider both current and future resource availability. Experimental results demonstrate that the proposed method not only sustains high acceptance ratios for resource-intensive services such as Cloud Gaming and VoIP but also significantly improves acceptance ratios for latency-critical categories such as Augmented Reality increases from 30\% to 50\%, while Industry 4.0 improves from 30\% to 45\%. Consequently, the prediction-based model achieves significantly lower E2E latencies of 20.5\%, 23.8\%, and 34.8\% reductions for VoIP, Video Streaming, and Cloud Gaming, respectively. This strategy ensures more balanced resource allocation, and reduces contention. 
\end{abstract}


%
\IEEEpeerreviewmaketitle

\section{Introduction}

In today's large-scale Data Centers (DC), Service Function Chaining (SFC) leverages the centralized programmability of Software-Defined Networking (SDN) and the on-demand instantiation of Network Function Virtualization (NFV) to interconnect virtual network functions (VNF) such as firewalls, Network Address Translation (NAT), into a coherent service pipeline \cite{intro1}. However, the resource consumption of each VNF fluctuates continuously with varying traffic loads and application demands \cite{intro2}. Relying on static resource allocations therefore risks either overprovisioning, wasting valuable resources, or underprovisioning. Addressing this challenge requires embedding accurate, advance forecasts of available computational power and storage directly into the process. By predicting resource demands ahead of time, the system can perform strategic placement of VNFs and DC selections, thereby ensuring Service Level Agreements (SLA) are met while making the most efficient use of available resources.

To address this challenge, recent studies have turned to Machine Learning (ML) for precise, advance forecasting of  resource usage. Techniques range from statistical time-series models (e.g., ARIMA) and ensemble learners (e.g., random forests) to Deep Learning (DL), such as Long Short Term Memory (LSTM) and Graph Neural Networks (GNNs) \cite{intro3}. By integrating these ML-derived predictions into SFC provisiong, systems can proactively optimize VNFs placement, and adaptively select DC. In practice, an ML-powered prediction system first ingests historical telemetry, such as resource utilization in past steps and are trained to minimize forecasting error (e.g., Mean Square Error), while predicting future step.

In parallel, Deep Reinforcement Learning (DRL) has been widely investigated for VNF placement and SFC provisioning. Unlike forecasting-based approaches that primarily focus on predicting future resource availability, DRL formulates the placement problem as a sequential decision-making task, where an agent learns optimal allocation strategies through interaction with the environment \cite{intro4}. By modeling the dynamic state of DC resources and SFC requirements, DRL-based methods have demonstrated effectiveness in maximizing acceptance ratios, reducing end-to-end latency, and improving overall resource utilization \cite{intro5}. 

Building on our prior study of DRL-based VNF placement \cite{arda24}, this work advances the problem formulation by incorporating predictive models into the framework. We employ DRL to generate datasets that capture DC resource utilization and SFCs/VNFs counts, which are subsequently used to train  forecasting models capable of predicting future availability resources, including Temporal Convolutional Network (TCN), LSTM, Temporal Graph Neural Networks (TGNN), Spatio-Temporal Graph Neural Networks (STGNN). Additionally after hyperparameter tuning for each model, an ensemble model is constructed. The ensemble model is then integrated into the DC selection criteria, allowing placement decisions to account not only for the current state of the infrastructure but also for predicted resource availability. This hybrid approach reduces the likelihood of resource contention, and improves the overall efficiency of SFC provisioning.

The main contribution of this paper is two-fold: 1) We extend DRL-based VNF placement by incorporating predictive intelligence through a hybrid DRL–ML framework. Specifically, we use DRL-generated datasets to train an ensemble prediction model composed of TGNN, STGNN, and LSTM, which have demonstrated more promising results in prior simulations. The ensemble’s forecasts of future data center resource availability are then integrated into the DC selection process, enabling prediction-assisted placement decisions that account for both current and anticipated resource states. 2)  We leverage Optuna for systematic hyperparameter optimization and design an ensemble of the best-performing models, ensuring accurate and stable predictions that strengthen the DC selection process.

The remainder of the paper is organized as follows: Section II presents the literature review, Section III describes the system model, Section IV discusses and evaluates performance, and Section V concludes the paper.

\section{Related Work}
DRL has attracted significant attention for addressing SFC provisioning challenges. 
Tran et al. \cite{tran2024drlsfc} propose a DRL-based framework for joint VNF embedding and routing for SFCs in NFV-enabled networks. Their approach, implemented with both Deep Q-Learning (DQN) and Advantage Actor-Critic (A2C) algorithms, demonstrates performance close to near-optimal optimization but with significantly reduced execution time.
Extending the focus toward multi-objective optimization in MEC, Xing et al. \cite{xing2024ddrlste} introduce a distributed DRL architecture with a transformer-based spatio-temporal encoder (DDRL-STE). Their approach jointly considers latency and load balancing and leverages distributed agents with pre-training to capture dynamic user preferences. 

To address reliability concerns, Zeng et al. \cite{zeng2023ruledrl} propose RuleDRL, a hybrid scheme that combines DRL with rule-based heuristics to achieve cost-effective and reliable SFC provisioning. By incorporating delayed reward modeling with bounded approximations, RuleDRL effectively mitigates constraint violations. To cope with stringent delay constraints in SFC deployment, Tian et al. \cite{tian2024dtssfc} propose a two-stage DRL-based deployment framework (DTS-SFC). The first stage uses a graph-based resource aggregation routing method to identify feasible candidate paths within delay constraints, while the second stage employs a Proximal Policy Optimization (PPO) agent to allocate VNFs along those paths. Their results confirm superior acceptance ratios and resource utilization under latency-sensitive workloads when compared with greedy and DRL baselines.
Together, these works demonstrate the versatility of DRL in tackling different challenges of SFC provisioning. However, none of them incorporate predictive modeling to anticipate future network states, a gap that our work directly addresses.

Recent works have advanced VNF workload prediction in SFCs by combining graph learning with temporal models. Wu et al. \cite{wu2023} introduce the Granularity-captured Graph Attention LSTM (GGAL), which leverages microservice-level granularity VNFs to better capture spatial correlations, coupled with LSTM for temporal dynamics. Their approach achieved state-of-the-art accuracy over LSTM, and graph-based baselines. Building on this, \cite{wu2024graph} propose a Graph Transformer with LSTM-attention for multi-step forecasting, enabling global structural modeling and long-term temporal dependency capture. This model outperformed classical and graph-based predictors. Together, these studies highlight the effectiveness of graph–temporal hybrids for fine-grained and forward-looking VNF load prediction in dynamic SFC environments. However, these works haven't investigated their integration with DRL agents or evaluated the impact on overall SFC provisioning and acceptance ratio.

\section{System Model}

\begin{figure*}[!hbt]
    \centering
    \includegraphics[width=0.61\textwidth, trim=0cm 0.0cm 0cm 0cm, clip]{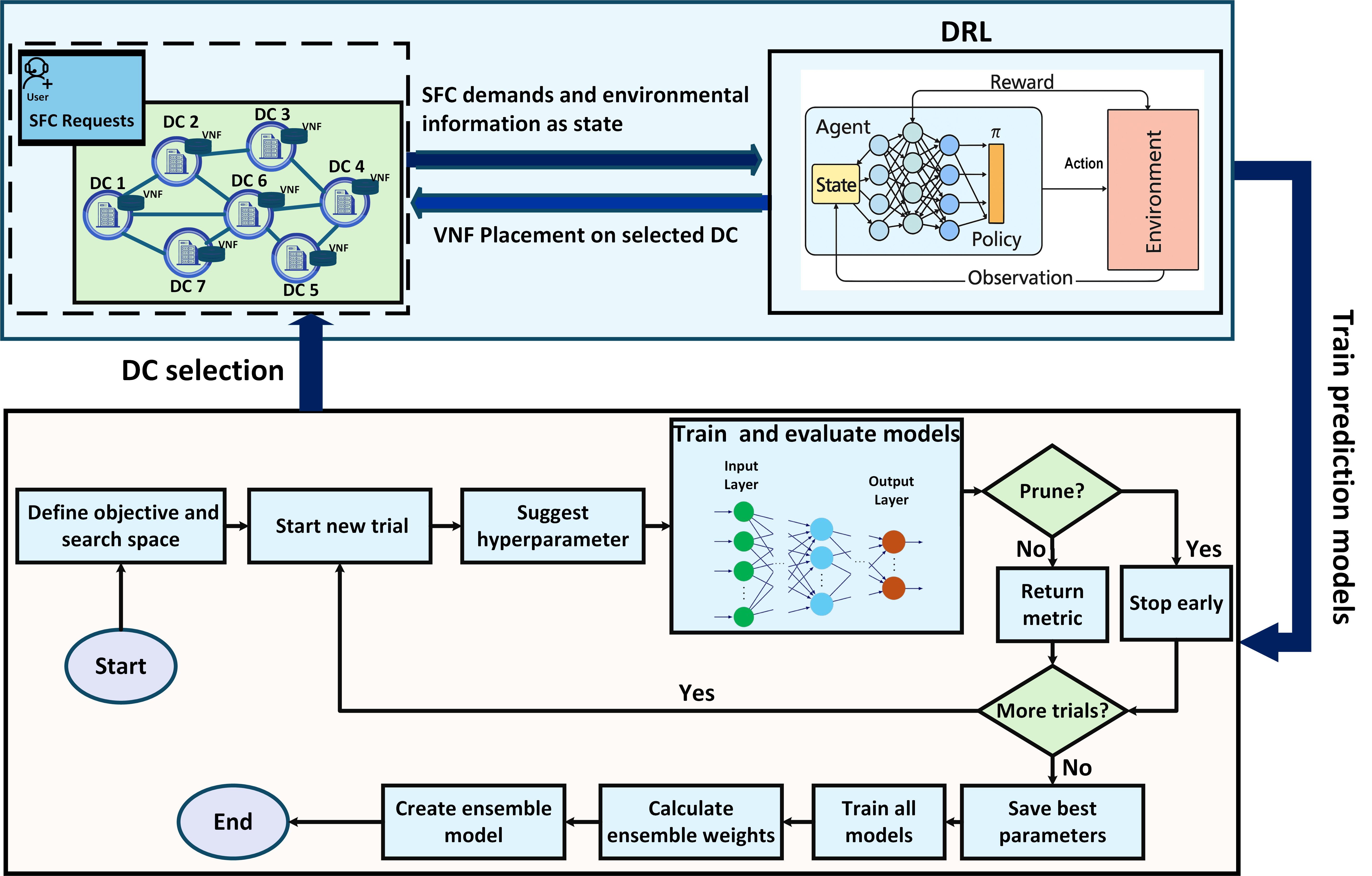}
\caption{Proactive SFC Provisioning with Forecast-Driven
DRL}
        \label{fig:framework}
\end{figure*}

As shown in \figurename \hspace{0.1pt}\ref{fig:framework} in our framework, RL is adopted to enable dynamic placement of incoming VNFs and SFCs. At each decision step, the RL agent observes the current network state, including available computational and storage, number of local and global VNFs, and number of local and global SFCs. The agent action corresponds to selecting a DC target to host the next VNF.
We generate a dataset from our DRL environment, following \cite{parisanof}. The dataset records, at each timestamp, the values of available storage, computational power, local SFC count, global SFC count, local VNF count, and global VNF count. We then train multiple prediction models on this dataset and employ Optuna for hyperparameter optimization. The best performing models are further combined into an ensemble predictor, which is subsequently integrated into DC selection. Prediction is not part of the DRL agent’s state and it is used to enable more efficient DC selection.

The DC selection process is structured into four categories. (i) DCs already hosting VNFs awaiting resources are prioritized to mitigate potential deadline violations. (ii) When sufficient historical state information exists, a prediction model (e.g., ensemble) forecasts upcoming computational and storage availability, and DCs with the highest predicted capacity are favored. If such data is unavailable, this step is skipped. (iii) DCs located along the shortest path between the current and destination DCs are selected to support low-latency routing. (iv) Remaining DCs serve as fallback options. By incorporating predictive modeling, the framework not only reacts to the current network state but also anticipates future resource availability, improving acceptance ratios in dynamic environments with fluctuating workloads. 

 \subsection{Problem Formulation}

The physical infrastructure is modeled as a graph $\mathcal{G} = (\mathcal{D}, \mathcal{L})$, where $\mathcal{D}$ is the set of VNFI-enabled DCs and $\mathcal{L}$ is the set of logical links. Each link $(i, j) \in \mathcal{L}$ has bandwidth capacity $B_{ij}$ (Mbps) and propagation delay $\tau_{ij}$, which is computed from the physical distance $\ell_{ij}$ divided by the propagation speed $c$. Each DC $i \in \mathcal{D}$ provides CPU capacity $\mathcal{C}_i$ (cycles/s) and storage capacity $\mathcal{M}_i$ (GB) \cite{arda24}. 
The set of VNFs is denoted by $\mathcal{V}$, where each type $v \in \mathcal{V}$ requires $\kappa^v$ CPU cycles and $\sigma^v$ storage for deployment. The supported service types set $\mathcal{S}$ includes Cloud Gaming (CG), Augmented Reality (AR), Video Streaming (VS), VoIP, Massive IoT (MIoT), and Industry~4.0 (In4). For each $s \in \mathcal{S}$, the corresponding SFC is an ordered sequence $\mathcal{F}^s = (f_1^s, \ldots, f_{N_s}^s)$ of VNFs from $\mathcal{V}$. Each service type $s$ is characterized by the tuple $\{\mathcal{F}^s, b^s, \Theta^s, \Lambda_s\}$, where $b^s$ is the bandwidth requirement, $\Theta^s$ is the maximum end-to-end (E2E) delay tolerance, and $\Lambda_s$ is the bundle size.
The objective is to maximize the acceptance ratio $A_R$, defined as the ratio between the total number of admitted requests $R_s$ and the total number of arrivals $\Lambda_s$. The optimization problem can be formulated as follows:

\begin{subequations}\small
\label{eq:opt_problem}
\begin{align}
& \underset{\mathbf{x},\mathbf{y}}{\text{maximize}} \quad
A_R = {\sum_{s \in \mathcal{S}} R_s}/{\sum_{s \in \mathcal{S}} \Lambda_s}
\label{eq:opt_problem_obj} \\[0.6em]
\text{s.t.} \quad
& \sum_{v \in \mathcal{V}} x_{i}^{v}\kappa^{v} \leq \mathcal{C}_{i},
&& \forall i \in \mathcal{D}, \label{eq:opt_problem_a} \\
& \sum_{v \in \mathcal{V}} x_{i}^{v}\sigma^{v} \leq \mathcal{M}_{i},
&& \forall i \in \mathcal{D}, \label{eq:opt_problem_b} \\
& \sum_{i \in \mathcal{D}} y_{i}^{f_k^s} = 1,
&& \forall s \in \mathcal{S}, \forall k, \label{eq:opt_problem_c} \\
& y_{i}^{f_k^s} \leq x_{i}^{f_k^s},
&& \forall s,k,i, \label{eq:opt_problem_d} \\
& \sum_{s \in \mathcal{S}} \sum_{k=1}^{N_s-1}
y_{i}^{f_k^s} y_{j}^{f_{k+1}^s} b^{s} \leq B_{ij},
&& \forall i\neq j, \label{eq:opt_problem_e} \\
& d_{\mathrm{prop}}^{s} + d_{\mathrm{proc}}^{s} \leq \theta^{s},
&& \forall s \in \mathcal{S}, \label{eq:opt_problem_f} \\
& x_i^{v}\in\mathbb{Z}_{\ge 0}, \;
y_i^{f_k^s}\in\{0,1\},
&& \forall i,v,s,k  \label{eq:opt_problem_g}
\end{align}
\end{subequations}

\begin{itemize}
    \item $\mathcal{S}$: set of service request types, with $\Lambda_s$ arrivals of type $s$ and $R_s$ admitted requests.
    \item $\mathcal{D}$: set of DCs; $\mathcal{V}$: set of VNFs.
    \item $x_i^v$: number of instances of VNF $v$ placed in DC $i$.
    \item $\kappa^v, \sigma^v$: CPU and storage requirement of VNF $v$.
    \item $\mathcal{C}_i, \mathcal{M}_i$: CPU and storage capacity of DC $i$.
    \item $y_i^{f_k^s}$: binary variable, $1$ if the $k$-th VNF of request $s$ is mapped to DC $i$, $0$ otherwise.
    \item $b^s$: bandwidth demand of request $s$
    \item $B_{ij}$: link capacity between DCs $i$ and $j$.
    \item $d_{\mathrm{prop}}^{s}$: propagation delay of request $s$, defined as
    \[\tiny
    d_{\mathrm{prop}}^{s} = \sum_{k=1}^{N_s-1} \sum_{i \in \mathcal{D}}\sum_{j \in \mathcal{D}} 
    y_{i}^{f_k^s} y_{j}^{f_{k+1}^s} \,\tau_{ij},
    \]
    where $\tau_{ij}=\ell_{ij}/c$, and $\ell_{ij}$ is the fiber length, and $c \approx 3\times 10^8$ m/s.
    \item $d_{\mathrm{proc}}^{s}$: processing delay of request $s$, given by
    \[\tiny
    d_{\mathrm{proc}}^{s} = \sum_{k=1}^{N_s} \sum_{i \in \mathcal{D}} 
    y_{i}^{f_k^s}\left(w_{i}^{f_k^s} + \rho_{i}^{f_k^s}\right),
    \]
    with $w_{i}^{f_k^s}$ the waiting time and $\rho_{i}^{f_k^s}$ the execution time.
    \item $\theta^s$: Max tolerable end-to-end delay for request $s$.
\end{itemize}

  \eqref{eq:opt_problem_a} ensures that the total CPU consumption of all VNFs deployed in a DC does not exceed its available computational capacity $\mathcal{C}_i$. Similarly,  \eqref{eq:opt_problem_b} limits the storage requirements of deployed VNFs so that they remain within the memory capacity $\mathcal{M}_i$ of each DC.  \eqref{eq:opt_problem_c} guarantees that every VNF in each service request is assigned to exactly one DC, which prevents both duplication and omission.  \eqref{eq:opt_problem_d} establishes the consistency between deployment and assignment by requiring that a VNF can only be allocated to a service request in a given DC if that VNF is actually instantiated there.  \eqref{eq:opt_problem_e} enforces the bandwidth limitation on logical links by ensuring that the aggregate traffic between consecutive VNFs of all SFCs does not exceed the link capacity $B_{ij}$.  \eqref{eq:opt_problem_f} enforces the service-level delay requirement: for each request, the total delay, consisting of propagation delay and processing delay, must not exceed the tolerated end-to-end latency $\theta^s$. Finally,  \eqref{eq:opt_problem_g} specifies the nature of the decision variables, where $x_i^v$ are non-negative integers representing the number of VNF instances deployed and $y_i^{f_k^s}$ are binary variables indicating whether a particular VNF in a service request is placed at a specific DC.

\textit{The state space} is composed of three feature groups: (i) DC–level information, including the number of installed and available VNFs, as well as remaining storage and CPU capacity; (ii) the local progress of SFCs, represented by the allocated and pending VNFs for each request type; and (iii) a global demand view, which captures the number of pending requests, their latency budgets, bandwidth requirements, and the VNFs still awaiting placement.
\textit{The action space} allows the agent to allocate a VNF, uninstall an existing VNF, or remain idle (Wait).
 \textit{The rewards} are selected based on the relative criticality of different outcomes in SFC provisioning. Successfully admitting and serving a request yields the highest reward (+2.0), reflecting the priority of maximizing acceptance. Dropped requests are penalized more heavily (-1.5) than invalid actions (-1.0) or unnecessary VNF uninstallations (-0.5), since they directly violate service-level objectives. The penalty magnitudes are calibrated such that the agent learns to avoid harmful behaviors while still being encouraged to explore feasible allocation strategies. Neutral outcomes are given zero reward to balance exploration and exploitation.

\subsection{Hyperparameter optimization}
We employ Optuna \cite{optuna1}, an automatic hyperparameter optimization framework utilizing Tree-structured Parzen Estimator (TPE) algorithms, to systematically identify optimal parameter configurations for neural network architectures \cite{optuna2}. 
Optuna defines an objective function (e.g. minimize error) that evaluates model performance, samples candidate hyperparameters in each trial, and prunes underperforming trials based on intermediate validation results to reduce computational overhead. By iteratively refining its search strategy, the framework converges toward the most effective configurations. In our case, LSTM, TCN, Temporal GNN, and Spatial-Temporal GNN models are optimized with the goal of minimizing prediction error on unseen data. Then, an ensemble model is built from the final models to improve robustness and accuracy on unseen data.


\subsection{Ensemble Model Formulation}

To take advantage of the complementary strengths of recurrent and graph-based architectures, 
we construct an ensemble model by combining the LSTM, TGNN, and STGNN forecasts, as these three show superior performance. This ensemble adaptively assigns weights to each base model, with the weights derived  from validation performance. To this end, we compute the normalized Root Mean Squared Error (nRMSE) for each feature across all 
validation windows obtained from the rolling splits. The nRMSE provides a scale-invariant measure of error by normalizing the RMSE with respect to the dynamic range of each feature. For model $m \in \{L,T,S\}$ (LSTM, TGNN, STGNN), feature $j$, and the set of validation time indices in fold $k$ ,$\mathcal{V}_k$, the normalized RMSE is defined as \eqref{eq:nrmse}. This normalization ensures that features with larger numerical ranges do not dominate the evaluation, allowing errors to be compared fairly across heterogeneous targets. To obtain a single performance score per model–feature pair, we aggregate all folds into a global nRMSE, which pools squared errors over every validation sample.
\begin{equation}\small
\mathrm{nRMSE}_{m}^{(j)} 
= \sqrt{\frac{1}{\sum_{k=1}^{K} |\mathcal{V}_k|} 
\sum_{k=1}^{K} \sum_{v \in \mathcal{V}_k} 
\left( \frac{ y_{v}^{(j)} - \hat{y}_{m,v}^{(j)} }
{ R_{k}^{(j)} } \right)^{2}}
\label{eq:nrmse}
\end{equation}
where $y_{v}^{(j)}$ is the ground truth, $\hat{y}_{m,v}^{(j)}$ is the prediction from model $m$, and 
$R_{k}^{(j)} = \max_{v \in \mathcal{V}_k} y_{v}^{(j)} - \min_{v \in \mathcal{V}_k} y_{v}^{(j)}$ is the dynamic range of feature $j$ \emph{within} fold $k$. 
To convert these validation errors into weights,  each model $m$ is assigned a score for each feature $j$ according to \eqref{eq:scores}:

\begin{equation}\small
s_{m}^{(j)} = - \big( \mathrm{nRMSE}_{m}^{(j)} \big)^{2}
\qquad m \in \{L,T,S\}
\label{eq:scores}
\end{equation}
A lower nRMSE corresponds to a higher score after negation and squaring, thereby penalizing larger 
errors more strongly. These scores are then normalized through a softmax transformation as \eqref{eq:softmax_weights}, which ensures that weights are non-negative (formulated in \eqref{eq:ensemble_feature_c}), sum to one (formulated in \eqref{eq:ensemble_feature_b}), and favor the model with lowest validation error. 
\begin{subequations}
\begin{equation}\small
w_{m}^{(j)} 
= \frac{\exp(s_{m}^{(j)})}{\exp(s_{\text{L}}^{(j)}) + \exp(s_{\text{T}}^{(j)}) + \exp(s_{\text{S}}^{(j)})}
\qquad m \in \{\text{L}, \text{T}, \text{S}\}
\label{eq:softmax_weights}
\end{equation}
\begin{equation}\small
w_{\text{L}}^{(j)}+ w_{\text{T}}^{(j)} + w_{\text{S}}^{(j)}= 1 \label{eq:ensemble_feature_b}
\end{equation}
\begin{equation}\small
w_{\text{L}}^{(j)}, \, w_{\text{T}}^{(j)}, \, w_{\text{S}}^{(j)} \geq 0
\label{eq:ensemble_feature_c}
\end{equation}
\end{subequations}

For each feature $j$, the prediction of ensemble model is formulated in in (\ref{eq:final_ensemble}), where weights are estimated according to \eqref{eq:softmax_weights}, and $\hat{y}_t^{(L,j)}$, $\hat{y}_t^{(T,j)}$, and $\hat{y}_t^{(S,j)}$  are the corresponding predictions from LSTM, TGNN, STGNN models, respectively. 

\begin{equation}\small
\hat{y}_{t}^{(j)} 
= w_{\text{L}}^{(j)} \, \hat{y}_{t}^{(\text{L},j)}
+ w_{\text{T}}^{(j)} \, \hat{y}_{t}^{(\text{T},j)}
+ w_{\text{S}}^{(j)} \, \hat{y}_{t}^{(\text{S},j)} 
\label{eq:final_ensemble}
\end{equation}


\begin{figure*}[htbp]
    \centering
    \begin{subfigure}{0.23\textwidth}
        \centering
        \includegraphics[width=\linewidth]{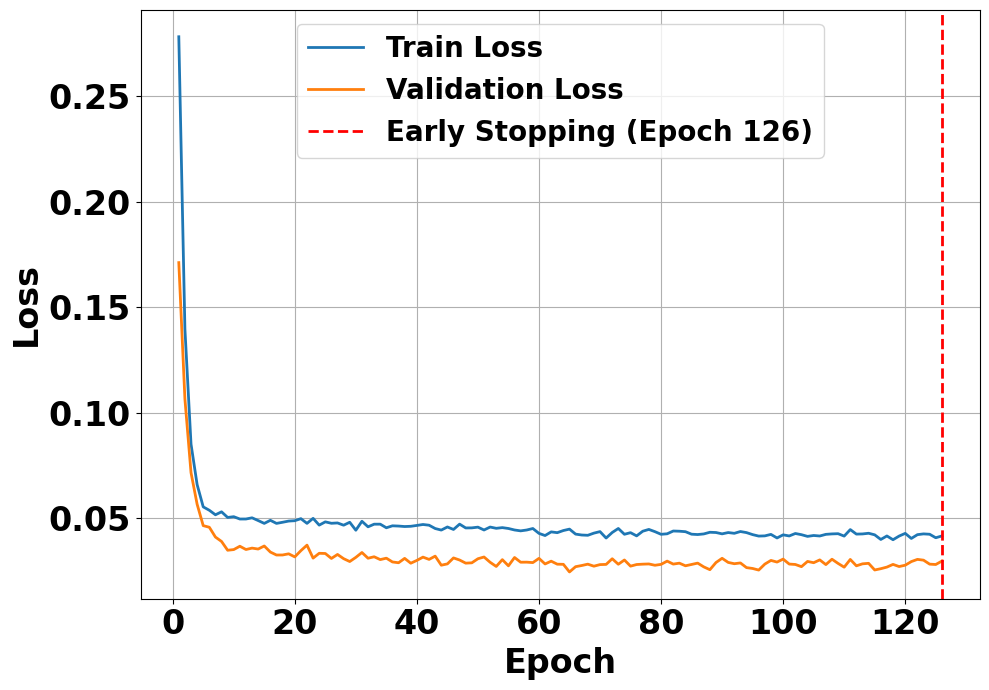}
        \caption{TCN }
        \label{fig:tcn_loss}
    \end{subfigure}
    \hspace{0.005\textwidth} 
    \begin{subfigure}{0.23\textwidth}
        \centering
        \includegraphics[width=\linewidth]{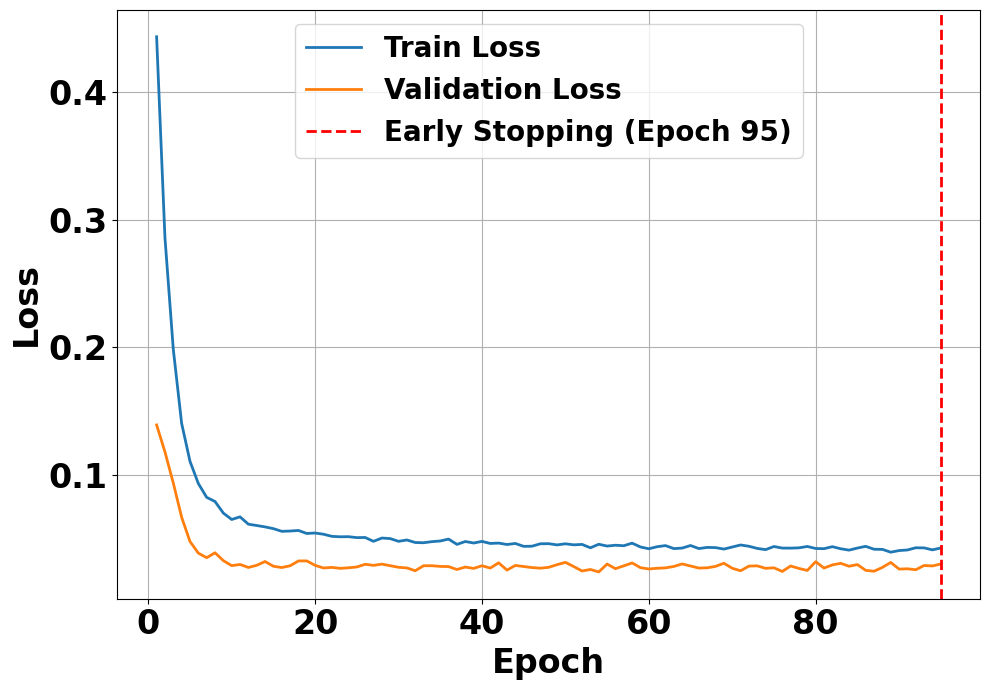}
        \caption{LSTM }
        \label{fig:lstm_loss}
    \end{subfigure}
      \hspace{0.005\textwidth} 
    \begin{subfigure}{0.23\textwidth}
        \centering
        \includegraphics[width=\linewidth]{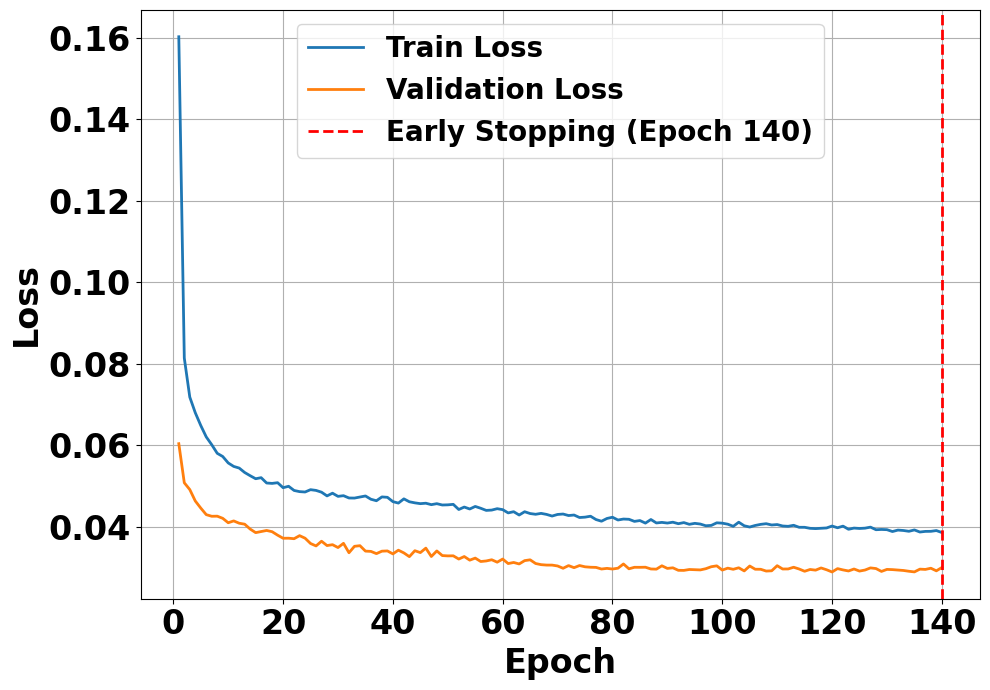}
        \caption{STGNN}
        \label{fig:stgnn_loss}
    \end{subfigure}
    \hspace{0.005\textwidth}
    \begin{subfigure}{0.23\textwidth}
        \centering
        \includegraphics[width=\linewidth]{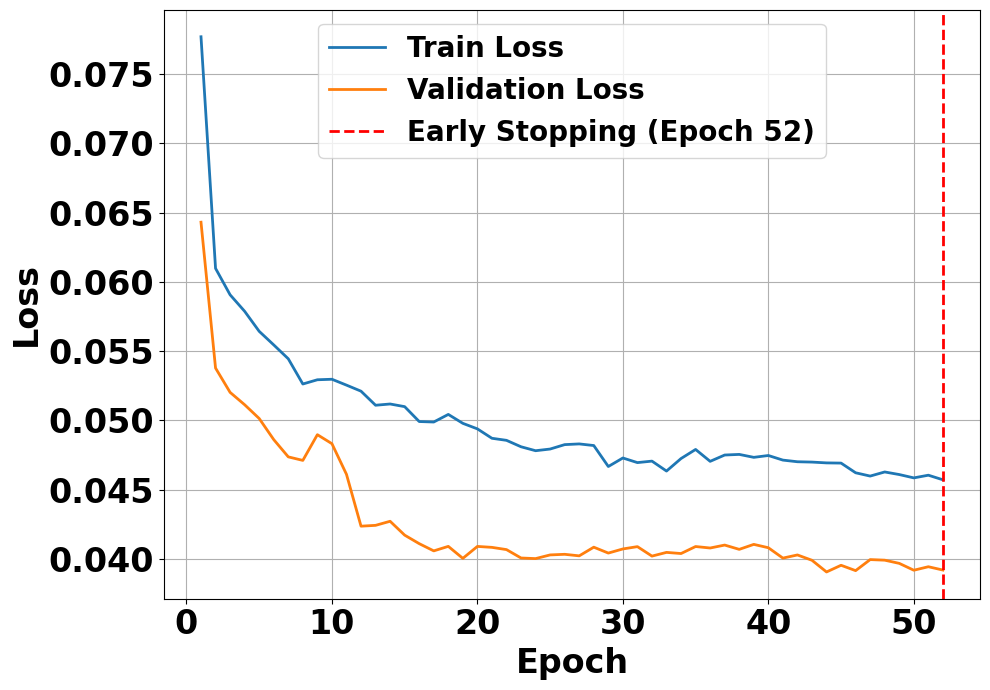}
        \caption{TGNN}
        \label{fig:tgnn_loss}
    \end{subfigure}
    \caption{Training and validation loss curves for the four models: TCN, LSTM, STGNN, and TGNN.}
    \label{fig:loss_curves}
\end{figure*}

\section{Performance Analysis}
The experiments were carried out on a system equipped with two NVIDIA NVIDIA GeForce RTX 3070. The simulation parameters and DRL parameters are set according to \cite{arda24}. We employ Optuna for hyperparameter optimization for prediction models, designing the search space by combining widely used defaults from the literature  as shown Table~\ref{tab:optuna_config}. To prevent overfitting, we employ early stopping, dropout, and weight decay/L2 regularization. We reserve the final 10\% of the series as a held-out test set. On the remaining data, we perform 5-fold rolling validation: in each fold, the training span grows forward in time and is followed by  10\% validation slice, ensuring robustness across different temporal segments while avoiding look-ahead bias. We construct input sequences with a window size of 20 and the window moves forward one time step each shift. At every shift, the model uses the previous 20 observations to predict the next time step, which encourages learning of sequential dependencies while limiting memorization and improving generalization.

The loss curves in Fig.~\ref{fig:loss_curves} demonstrate the training dynamics of the LSTM, STGNN, TCN, and TGNN models. Across all four architectures, the training loss decreases rapidly during the early epochs and stabilizes afterward, while the validation loss follows a similar trend with close alignment to training loss. This behavior indicates that the models achieve good generalization without significant overfitting. The early stopping prevents unnecessary training beyond the plateau phase, ensuring computational efficiency while preserving performance. Notably, the STGNN achieves the lowest validation loss overall, suggesting its superiority.

Table~\ref{tab:nrmse} reports nRMSE of all models across six features,. The ensemble model is constructed by combining LSTM, TGNN, and STGNN predictions, with weights derived from their best hyperparameter-tuned variants identified through Optuna. This ensures that the ensemble leverages the complementary strengths of individually optimized models. As shown in Table~\ref{tab:nrmse}, STGNN achieves the lowest errors on two metrics (computational power and global SFC), confirming its strength in modeling joint temporal–spatial dependencies. By contrast, the ensemble outperforms STGNN on the remaining four features (storage, global VNF, local VNF, local SFC count), highlighting the benefit of combining models. TCN consistently shows the highest errors. We also compute the winner rate, which measures how often a model achieves the lowest error across different features and DCs. While nRMSE summarizes the overall error, winner rate highlights consistency, as a model may not always have the best average score, but if it frequently wins on different cases, it indicates stronger robustness. The ensemble achieves the highest winner rate (40\%), meaning it most frequently outperforms all other models. STGNN follows with 23\%, demonstrating strong but less consistent dominance, while TGNN (22\%) and LSTM (11\%) contribute in more specific scenarios. TCN only reaches 4\%, reinforcing its limited suitability for this workload. Together, these results show that the ensemble model is the most robust choice overall.

\begin{table}[!t]
\centering
\caption{Hyperparameters and Training Configuration}
\label{tab:optuna_config}
\begin{tabular}{|l|p{4.2cm}|}
\hline
\textbf{Hyperparameter} & \textbf{Search Space / Value} \\
\hline

Learning rate ($lr$)  & $[10^{-4}, 10^{-2}]$ (log-uniform )\\
Weight decay          & $[10^{-6}, 10^{-2}]$ ( log-uniform)  \\
Batch size  & \{8, 16, 32, 64\}  \\
Hidden units (LSTM)   &  \{16, 32, 64, 128\}   \\
Number of  layers & \{1, 2, 3\}  \\
Dropout        & \{0.1- 0.5\}  (step size 0.1) \\
Channels (TCN)        & \{16, 32, 64\}  \\
Hidden size (GNN)     & \{16, 32, 64, 128\} \\
Max epochs  & 200  \\
Early stopping patience & 15 epochs with min $\Delta = 10^{-6}$ \\
Minimum epochs& 20  \\
Number of trials & 50  \\
\hline
\end{tabular}
\end{table}

\begin{table}
\centering
\caption{Average nRMSE per feature across all models}
\label{tab:nrmse}
\resizebox{\columnwidth}{!}{%
\begin{tabular}{lccccc}
\hline
\textbf{Metric} & \textbf{LSTM} & \textbf{TCN} & \textbf{TGNN} & \textbf{STGNN} & \textbf{Ensemble} \\
\hline
Available computational power & 0.184 & 0.212 & 0.162 & \textbf{0.137} & 0.142 \\
Available storage             & 0.153 & 0.248 & 0.172 & 0.157 & \textbf{0.141} \\
Global sfc count              & 0.171 & 0.195 & 0.155 & \textbf{0.129} & 0.138 \\
Global vnf count              & 0.178 & 0.257 & 0.141 & 0.146 & \textbf{0.133} \\
Local sfc count               & 0.191 & 0.235 & 0.168 & 0.146 & \textbf{0.137} \\
Local vnf count               & 0.191 & 0.276 & 0.155 & 0.142 & \textbf{0.140} \\
\hline
\end{tabular}%
}
\end{table}

\begin{figure*}
    \centering
        \includegraphics[width=0.99\linewidth, trim=1mm 8mm 2mm 0, clip]{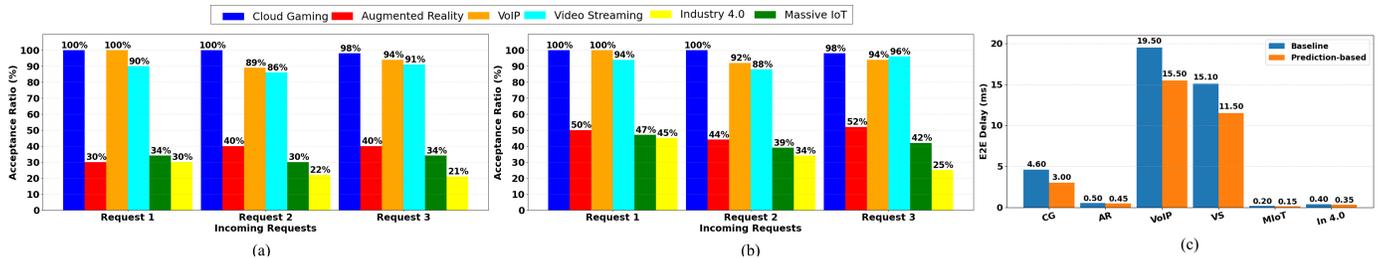}
    \caption{Acceptance Ratio for (a) baseline and (b) prediction-based model, and (c) E2E latencies for all SFC types }
    \label{fig:acceptance_ratio}
\end{figure*}

To evaluate the effectiveness of the proposed proactive Dc selection, we compare the acceptance ratio of SFC requests under two schemes: (i) a baseline model, where DC selection is performed without considering next step resource availability, and (ii) the proposed prediction-based Dc selection, where DC selection leverages predictive insights from ensemble for Claoud Gaming (CG), Augumented Reality (AR), Voice over IP (VoIP), Video Streaming (VS), Industry 4.0 (In4.0) and Massive IoT (MIoT).
 As shown in Fig.~\ref{fig:acceptance_ratio}, the baseline approach achieves high acceptance for resource-intensive services such as Cloud Gaming and VoIP (close to 100\%). However, the acceptance ratio for smaller but latency-critical categories, such as AR and In4.0, remains low (20–34\%). Similarly, MIoT requests also exhibit lower admission rates, reflecting the inability of the baseline scheme to balance allocations across diverse service classes. In contrast, the proposed prediction-based DC selection improves acceptance ratio across all under-served categories. For instance, AR requests increase from 30\% in the baseline to 45–50\% with prediction, while In4.0 improves from 21–30\% to 25–45\%. MIoT acceptance rises from 22–34\% to 30–47\%. Importantly, these improvements are achieved without compromising the already high acceptance rates of CG, VoIP, and VS, which remain above 90\%. 
The E2E latencies are calculated using only accepted SFC requests and shown in Fig.~\ref{fig:acceptance_ratio}-c. Since the E2E latency tolerances for AR, MIoT, and In4.0 services are inherently stringent, both the baseline and prediction models prioritize these SFCs to ensure compliance with their delay constraints. Consequently, their E2E delays remain comparable across the two models.
In contrast, the prediction-based model demonstrates a significant improvement for VoIP, VS, and CG, achieving approximately 20.5\%, 23.8\%, and 34.8\% lower E2E delays, respectively, compared to the baseline.

\section{Conclusion}
This paper presented a hybrid forecasting–DRL framework for proactive SFC provisioning. By combining DRL-generated datasets with ensemble forecasting models optimized via Optuna, our approach enables placement decisions that anticipate future resource demands rather than reacting solely to current states. The results confirm that this predictive strategy not only maintains strong acceptance ratios for resource-intensive services but also significantly improves acceptance ratio for latency-sensitive categories such as Augmented Reality and Industry 4.0 from 30\% to 50\% and from 30\% to 45\%, respectively. Moreover, this strategy effectively reduces E2E latency, achieving up to 34.8\%, 23.8\%, and 20.5\% delay reductions for Cloud Gaming, Video Streaming, and VoIP, respectively, compared to the baseline.
Looking ahead, incorporating more sophisticated spatio-temporal predictors, such as graph-based transformers, could enhance the ability to capture cross-DC interactions and long-range temporal patterns. 
 

\section*{Acknowledgment}
This work is supported by the Natural Sciences and Engineering
Research Council of Canada (NSERC) Alliance Program, MITACS
Accelerate Program, and NSERC CREATE TRAVERSAL program.
\bibliographystyle{IEEEtran}

\end{document}